\documentclass[12pt]{article}
\usepackage{jheppub}

\pdfoutput=1

\usepackage{amsmath,bbm,array,amsfonts,graphicx,wrapfig,lscape,float,mathtools,multirow,longtable}
\usepackage[dvipsnames]{xcolor}
\usepackage{array}

\newcommand{\be}{\begin{equation}}
\newcommand{\ee}{\end{equation}}
\newcommand{\beq}{\begin{equation}}
\newcommand{\beql}[1]{\begin{equation}\label{#1}}
\newcommand{\eeq}{\end{equation}}
\newcommand{\ba}{\begin{array}}
\newcommand{\ea}{\end{array}}
\newcommand{\bea}{\begin{eqnarray}}
\newcommand{\beal}[1]{\begin{eqnarray}\label{#1}}
\newcommand{\eea}{\end{eqnarray}}
\newcommand{\ben}{\begin{enumerate}}
\newcommand{\een}{\end{enumerate}}
\newcommand{\bean}{\begin{eqnarray*}}
\newcommand{\eean}{\end{eqnarray*}}
\newcommand{\eref}[1]{(\ref{#1})}
\newcommand{\sref}[1]{\S\ref{#1}}

\newcommand{\fref}[1]{Figure \ref{#1}}
\newcommand{\btab}[1]{\begin{tabular}{#1}}
\newcommand{\etab}{\end{tabular}}

\def \Black {\pdfsetcolor{0 0 0 1}}

\newcommand{\comment}[1]{}

\newcommand{\qed}{\nobreak \ifvmode \relax \else
      \ifdim\lastskip<1.5em \hskip-\lastskip
      \hskip1.5em plus0em minus0.5em \fi \nobreak
      \vrule height0.75em width0.5em depth0.25em\fi}

\definecolor{darkspringgreen}{rgb}{0.09, 0.45, 0.27}
\definecolor{forestgreen}{rgb}{0.13, 0.55, 0.13}

\usepackage{array}
\usepackage{comment}
\usepackage{physics}
\usepackage{subcaption}
\usepackage{tikz,tikz-3dplot}
\usepackage[colorlinks=true]{hyperref}
\newcolumntype{C}[1]{>{\centering\let\newline\\\arraybackslash\hspace{0pt}}m{#1}}

\definecolor{yellow2}{rgb}{0.98, 0.80, 0.20}

\title{The 5d Tangram: Brane Webs, 7-Branes and Primitive T-Cones}

\author[a,b]{Ignacio Carre\~no Bolla,}

\author[c,d,e]{Sebasti\'an Franco,}

\author[a,b]{Diego Rodr\'iguez-G\'omez}

\affiliation[a]{Department of Physics, Universidad de Oviedo \\  
C/ Federico Garc\'ia Lorca  18, 33007  Oviedo, Spain}
\affiliation[b]{Instituto Universitario de Ciencias y Tecnolog\'ias Espaciales de Asturias (ICTEA) \\
 C/~de la Independencia 13, 33004 Oviedo, Spain.}
\affiliation[c]{Physics Department, The City College of the CUNY\\
	160 Convent Avenue, New York, NY 10031, USA}
\affiliation[d]{Physics Program and \textsuperscript{$e$}Initiative for the Theoretical Sciences\\
	The Graduate School and University Center, The City University of New York\\
	365 Fifth Avenue, New York NY 10016, USA}

\emailAdd{ignaciocarbolla@gmail.com}	
\emailAdd{sfranco@ccny.cuny.edu}
\emailAdd{d.rodriguez.gomez@uniovi.es}

\abstract{Two highly successful approaches to constructing 5d SCFTs are geometric engineering using M-theory on a Calabi-Yau 3-fold and the use of 5-brane webs suspended from 7-branes in Type IIB string theory. In the brane web realization, the extended Coulomb branch of the 5d SCFT can be studied by opening the web using rigid triple intersections of branes--i.e. configurations with no deformations. In this paper, we argue that the geometric engineering counterpart of these rigid triple intersections are the T-cones introduced in the mathematical literature. We extend the class of rigid brane webs to include locked superpositions of the minimal ones. These rigid brane webs serve as fundamental building blocks for supersymmetrically tessellating Generalized Toric Polygons (GTPs) from first principles. Interestingly, we find that the extended Coulomb branch generally exhibits a structure consisting of multiple cones intersecting at a single point. Hanany-Witten (HW) transitions in the web have been conjectured to correspond geometrically to flat fibrations over a line, where the central and generic fibers represent the geometries dual to the webs before and after the transition. We demonstrate this explicitly in an example, showing that for GTPs reducing to standard toric diagrams, the HW transition corresponds to a deformation of the BPS quiver that we map to the geometric deformation.}

\begin{document}

\maketitle

\section{Introduction and summary}

String/M Theory provides various approaches to engineer $5d$ Superconformal Field Theories (SCFTs). They can be geometrically realized by M-theory on a local Calabi-Yau 3-fold (CY$_3$) \cite{Morrison:1996xf,Intriligator:1997pq}. The best studied examples of this approach involve toric CY$_3$'s.  Alternatively, $5d$ SCFTs can be realized on the worldvolume of webs of $(p,q)$ 5-branes in type IIB String Theory \cite{Aharony:1997ju, Aharony:1997bh}. These two constructions are elegantly related through dualities. Reducing on a $\mathbb{T}^2$ inside the toric fiber, M-theory gives rise to Type IIB 5-branes located at the locus where the $\mathbb{T}^2$ degenerates, giving rise to the corresponding 5-brane web. In other words, the $(p,q)$-brane can be regarded as the spine of the toric diagram for the CY$_3$ \cite{Aharony:1997bh,Leung:1997tw}.

Webs of $(p,q)$ 5-branes and the associated $5d$ theories can be extended by introducing 7-branes on which one or multiple branes 5-branes can terminate \cite{Aharony:1997ju}. Such brane configurations are captured by the so-called Generalized Toric Polygons (GTPs). In recent years, significant efforts have been devoted to understanding the $5d$ theories and geometries associated with GTPs (see, e.g., \cite{Kim:2014nqa,Bourget:2023wlb,Franco:2023flw,Franco:2023mkw,Cremonesi:2023psg,Arias-Tamargo:2024fjt}). This paper studies the physics and geometry of T-cones, a special class of triangular GTPs, which we propose should be viewed as fundamental building blocks of the theories associated with GTPs. 

When the CY$_3$ is a standard toric geometry, its geometric resolutions correspond to motion along the extended Coulomb branch of the $5d$ theory, while its deformations correspond to motion along the Higgs branch. In this work, we introduce an additional geometric operation frequently discussed in the mathematical literature: $\mathbb{Q}$-Gorenstein smoothing (often referred to as $\mathbb{Q}$G-smoothing, or simply smoothing). T-cones are special in that they represent minimal geometries that are  $\mathbb{Q}$G-smoothable.

Below we summarize some of our main findings.

\begin{itemize}
\item T-cones corresponds to special triple junctions of 5-branes suspended from three 7-branes with no extended Coulomb branch.
\item The smoothing of T-cones translates to a reduction of the triple junction of 5-branes, via brane crossing, into a configuration with 5-branes stretch between only two of the 7-branes, while the third 7-brane is detached.
\item T-cones do not possess an extended Coulomb branch or any continuous global symmetry. However, we argue that they are nonetheless non-trivial as $5d$ SCFTs. They have a $\mathbb{Z}_p$ 1-form symmetry and they also carry at least a $\mathbb{Z}_p$ gauge theory.

\end{itemize}

The extended Coulomb branch of a generic GTP can be understood through a combination of resolution and smoothing, represented by a polygonation of the GTPs, with T-cones playing a central role. This polygonation proceeds as follows: 

\begin{itemize}
\item As mentioned earlier, we regard T-cones as elementary building blocks in this tessellation of the GTP. We distinguish between primitive T-cones (which have two sides of lattice length 1) and non-primitive T-cones. Note that the elementary triangles in standard resolutions are a special case of primitive T-cones.

\item Given a GTP, we can tile it with T-cones, resulting in a smooth space made up of locally resolved and smoothed singularities glued together. The singular SCFT limit is reached by setting both the resolutions and smoothings to zero. This clearly demonstrates that the singular CY geometry associated with a GTP is the same as that of the underlying toric diagram.

\item However, primitive T-cones alone are generally insufficient for tessellating general GTPs, and additional building blocks are required.

\item The most general building blocks for GTPs include not only primitive T-cones but also a new class of objects we refer to as {\it locked superpositions}, which can be understood as combinations primitive T-cones and other subwebs superposed in such away that the motions of their external legs are restricted. They correspond to pieces in the diagram not covered by primitive T-cones. However, their physical realization allows us to go further, providing insights into how they can be smoothed.
\item Since the different allowed building blocks have different areas, different possible tessellations generically show a to different number of faces. This suggests that the extended Coulomb branch is generically composed of the union of different cones.
\end{itemize}

The $\mathbb{Q}$G-smoothing is a particular instance of brane crossing. More generically, it was conjectured in \cite{Arias-Tamargo:2024fjt} that Hanany-Witten transitions whereby an external 7-brane is crossed to the other side of the web (and possibly extra five-branes are created/destroyed) geometrically correspond to the existence \cite{Ilten_2012} of a flat fibration over $\mathbb{P}^1$ where the central fiber corresponds to the geometry dual to the original web and the non-central fibers to the geometry dual to the crossed web. At the level of the BPS quiver, it was conjectured that this corresponds to the deformation introduced in \cite{Cremonesi:2023psg}. We provide an explicit example demonstrating this claim.

\paragraph{Note added:} As we were finalizing this note, \cite{AAB} appeared on the arXiv, showing some overlap with our work. Our results appear to be consistent with those presented in \cite{AAB}.

\section{Primitive T-cones and 7-Branes}

Brane webs engineering 5d SCFT's are naturally regarded as suspended from 7-branes whose branch cuts are oriented radially outgoing. As the position of the 7-branes along the corresponding legs are not parameters of the 5d SCFT, these can be crossed to the other side without altering the 5d SCFT. Then, when rotating the crossed 7-brane branch cut so that it becomes radially oriented after the crossing it sweeps half of the web changing the $(p,q)$ types of the legs. In turn, the crossing can create extra branes due to the Hanany-Witten effect (see the appendix in \cite{Bergman:2020myx} for a review of the crossing process). This brane crossing process produces a new and generically different looking web which is nevertheless described by the same 5d SCFT. Very recently, it has been suggested that the mathematical avatar of this process is the so-called mutation of the GTP associated to the brane web \cite{Franco:2023flw,Arias-Tamargo:2024fjt}. As described in \cite{Arias-Tamargo:2024fjt}, an essential ingredient in the mutation process is the so-called {\it primitive T-cone}, which corresponds to a triangle with one side along a side of the GTP and whose opposite vertex is strictly in the interior of the GTP (we will be more precise below) and  which is naturally regarded as an external leg, 7-brane included, of the brane web. Then, roughly speaking, the mutation amounts to erasing a primitive T-cone at one side of the GTP and pasting it to the other side so that the creation of branes in this case is encoded in the properties of the T-cone to glue, thus mimicking the brane crossing in Physics language. Moreover, a theorem by Ilten \cite{Ilten_2012} ensures that polytopes related by mutation can be regarded as fibers of a flat family over a line. A classical example is the mutation of $\mathbb{F}_2$ into $\mathbb{F}_0$, which in that case corresponds to the classical smoothing of the singularity. For toric GTPs, this was conjectured in \cite{Arias-Tamargo:2024fjt} to correspond to the deformation at the level of the BPS quiver introduced in \cite{Cremonesi:2023psg}. In section \ref{explicit} we will show an explicit example supporting this claim. Finally, it is important to stress that the notion of mutation introduced in the mathematical literature requires the apex of the primitive T-cone to be in the strict interior of the GTP, which then must contain internal dots. Yet, it is clear that the physical intuition suggests that this requirement could be dropped.

\begin{figure}[h!]
\centering
\includegraphics[height=5cm]{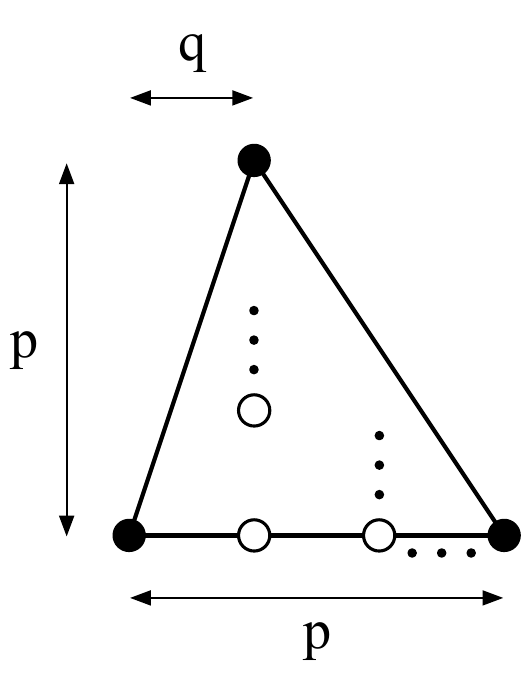}
\caption{GTP for a primitive T-cone.}
\label{GTP T-cone 1}
\end{figure}

At any rate, primitive T-cones\footnote{For brevity, we will often refer to them simply as T-cones.} appear as atomic ingredients when studying generic brane webs, which motivates their study by themselves. Originally, T-cones were introduced in \cite{Shepherd1988}, where it was argued that they correspond to minimal geometries that admit $\mathbb{Q}$-Gorenstein smoothing \cite{Shepherd1988}.\footnote{This means that there exists a flat family $\psi:\mathcal{X}\rightarrow \Delta$ where $\Delta$ is a small disk (that is, locally $\mathbb{C}$) such that $a)$ the general fiber $X_{t\ne 0}$ is smooth, $b)$ the central fiber is (isomorphic to) the singular variety --in this case $X_0\sim\mathbb{C}^3/\mathbb{Z}_{p^2}$--, $c)$ the canonical divisor $K_{\chi/\Delta}$ is $\mathbb{Q}$-Cartier.} 
They correspond to a class of triangular GTPs with the following properties:
\begin{itemize}
\item Only the corner dots are black.
\item The length of the base and the height of the triangle are equal. We will call this number $p$. The base of the triangle therefore contains $p-1$ white dots.
\item The two sides of the triangle other than the base do not contain internal points.
\end{itemize}
\fref{GTP T-cone 1} shows the general structure of a T-cone. Up to $SL(2,\mathbb{Z})$ transformations, its corners can be positioned at $(0,0)$, $(p,0)$ and $(q,p)$, where $p$ and $q$ are coprime. For a given $p$, there might be multiple inequivalent primitive T-cones, which would correspond to different positions of the top vertex in \fref{GTP T-cone 1}. We will return to this question below.

\begin{figure}[h!]
\centering
\includegraphics[height=5cm]{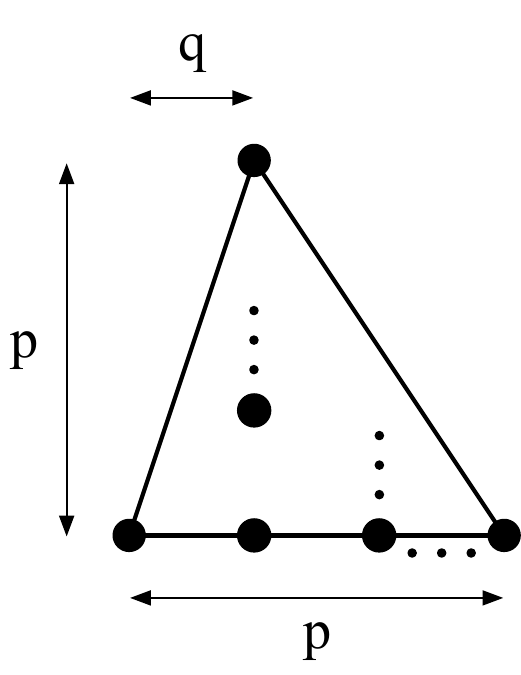}
\caption{Toric diagram obtained by turning all dots in the GTP of primitive T-cone black.}
\label{Toric T-cone 1}
\end{figure}

Considering the GTPs of T-cones as standard toric diagrams, as illustrated in \fref{Toric T-cone 1} (where all points are represented as black dots), we observe that they correspond to $\mathbb{C}^3/\mathbb{Z}_{p^2}$ orbifolds with a single generator acting as follows:
\beq
(z_1,\,z_2,\,z_3)\sim (\xi^p\,z_1,\,\xi^{p-q}\,z_2,\,\xi^{-2p+q}\,z_3)\,,\qquad \xi^{p^2}=1\,,
\eeq
where $q$ is coprime with $p$ for $p\geq 2$.\footnote{As orbifold actions, the inequivalent ones are actually reduced to $q\in [1,\lfloor\frac{p}{2}\rfloor]$.} For $p=1$ we have $q=0$, and the T-cone corresponds to the familiar minimal triangle of toric resolutions. The order $p^2$ of the orbifold follows from the area of the toric diagram in terms of minimal triangles. 

To compare with some results in the literature, we can take a different, yet equivalent, form of the orbifold action summarized in the vector $(a_1,a_2,a_3)$ with $a_1+a_2=p$, $a_1$ and $a_2$ coprime, and $a_1+a_2+a_3=p^2$. Interestingly, the number of T-cones for each value of $p$ is determined by Euler’s Half Totient function, revealing a fascinating link between number theory and the classification of $5d$ theories. The distinct orbifolds for $2\leq p \leq 5$ are:
\beq
\begin{array}{|c|c|c|}
\hline
\ \ p \ \ \ & \ \mbox{Orbifold of } \mathbb{C}^3 \ & \ \ (a_1,a_2,a_3) \ \ \ \ \\[.5mm] \hline 
2 & \mathbb{Z}_{4} & (1,1,2) \\[.5mm] \hline 
3 & \mathbb{Z}_{9} & (1,2,6) \\[.5mm] \hline 
4 & \mathbb{Z}_{16} & (1,3,12) \\[.5mm] \hline 
5 & \mathbb{Z}_{25} & (1,4,20) \\[.5mm] 
 & & (2,3,20) \\[.5mm] \hline 
\end{array}
\eeq
The $p=2$ and $3$ cases have appeared in the classification of orbifolds of \cite{Davey:2010px}. We observe that the first multiple solutions arise at $p=5$.

\subsection{The physics of T-cones} 

\begin{figure}[h!]
\centering
\includegraphics[height=5cm]{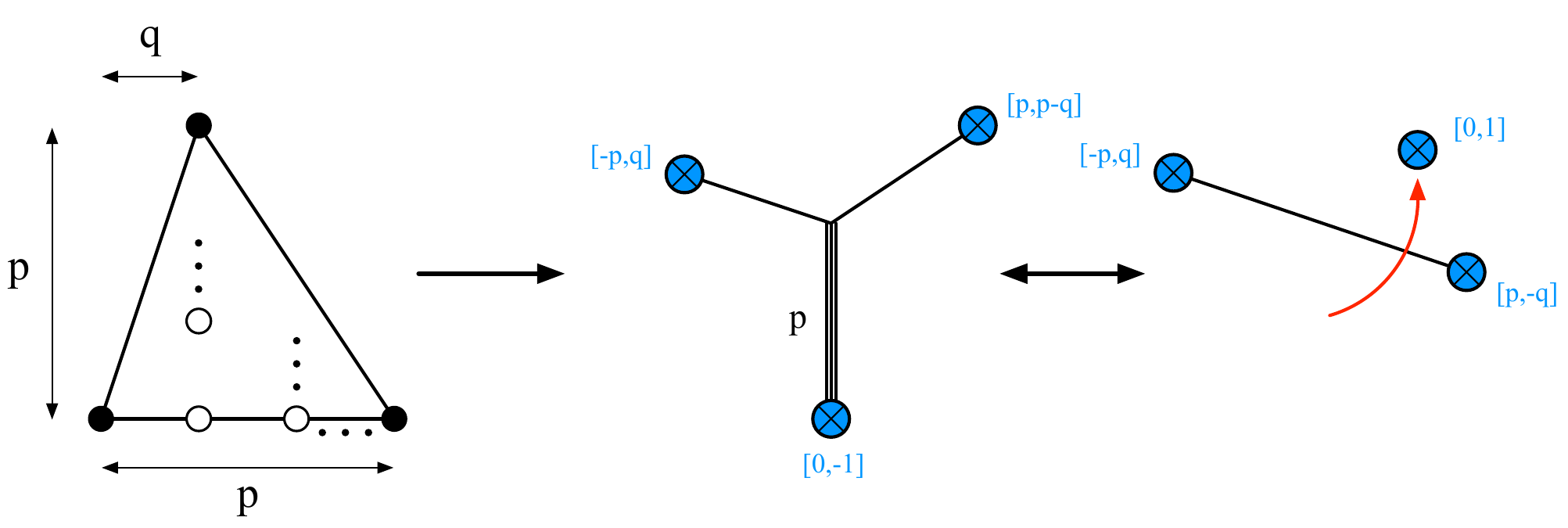}
\caption{A T-cone, its associated brane web and the detachment of one of the 7-branes.}
\label{GTP - Web - T-cone}
\end{figure}

Let us now discuss the physical interpretation of T-cones. T-cones correspond to Y-junctions of 5-branes suspended from three 7-branes, with one of the 7-branes attached to $p$ 5-branes and each of the other two 7-branes connected to a single 5-brane, as shown in \fref{GTP - Web - T-cone}. 

What makes this class of brane configurations special? The 7-branes can move freely along the corresponding legs, undergoing a Hanany-Witten (HW) transition when they cross the web. Moving the $[0,-1]$ 7-brane across the web, it turns into a completely detached $[0,1]$ 7-brane. We are left with a 5-brane stretched between a $[-p,q]$ and $[p,-q]$ 7-branes.
Alternatively, we can imagine starting from the configuration with the detached D7-brane and running the HW transition in the opposite direction to generate the Y-junction.  It is natural to identify this with the fact that T-cones are $\mathbb{Q}$G-smoothable \cite{Shepherd1988}: starting with the toric diagram corresponding to coloring in black all dots in the T-cone, the smoothing would be converting the T-cone into a GTP, in particular introducing one $(0,1)$ 7-brane on which $p$ NS5 branes end. The disk $\Delta$ parametrized by $t$ is then a proxy for the original GTP ($t=0$) or the crossed, detached, version $t\ne 0$. 

We propose that T-cones are also natural building blocks from a $5d$ perspective. Consider the task of characterizing $5d$ SCFTs that are as minimal as possible. By this, we mean theories with no extended Coulomb branch, namely with a 0-dimensional Coulomb branch and no mass deformations. For 5d SCFTs engineered on brane webs, this implies webs with three external legs (which implies that the theories have a rank 0 global symmetry) and self-intersection equal $-2$ \cite{Bergman:2020myx} (see appendix \sref{SI}). This is equivalent to demanding that upon HW crossing an external 7-brane, it emerges detached. Thus, all such minimal $5d$ SCFTs can be realized (up to $SL(2,\mathbb{Z})$ transformations) by taking a $(p,q)$ segment and crossing a $(0,1)$ 7-brane, for all possible non-equivalent choices of $(p,q)$. This is precisely the construction outlined in the previous paragraph and resulting in the Y-junctions in \fref{GTP - Web - T-cone}. 

Although T-cones, as 5d SCFTs, have no Coulomb branch or mass deformations, we argue that they are not trivial. The simplest example of T-cone with $(p,q)=(2,1)$ arises as the remnant after going far into the Higgs branch of the $E_1$ theory. The latter possesses a $\mathbb{Z}_2$ 1-form symmetry that cannot be broken by VEVs of local operators. Therefore, we expect the $(2,1)$ T-cone to correspond to a $5d$ SCFT with a $\mathbb{Z}_2$ 1-form symmetry.\footnote{It is natural to associate it to a $5d$ $\mathbb{Z}_2$ gauge theory: the $E_1$ theory admits a mass deformation into pure $SU(2)$ SYM. In this description, all matter --including instantons-- is in the adjoint representation of $SU(2)$, to which the $\mathbb{Z}_2$ center is insensitive. It is natural to conjecture that this piece is responsible for the $\mathbb{Z}_2$ 1-form symmetry even at infinite coupling.} More generally, combining the methods in \cite{Morrison:2020ool} with the assumption that the 1-form symmetry of a GTP equals that of the underlying toric diagram, we conjecture that a $(p,q)$ T-cone corresponds to a $5d$ SCFT with no Coulomb branch, no mass-deformations and $\mathbb{Z}_p$ 1-form symmetry.

\subsection{T-cones beyond primitivity} \label{noprim}

As mentioned earlier, the GTP corresponding to a primitive T-cone has vertices $(0,0)$, $(p,0)$ and $(q,p)$, with $p$ and $q$ are coprime. This last property ensures that except for the length $p$ base, the other two sides do not have internal points. Let us consider what happens if we do not require $p$ and $q$ to be coprime, but we still keep the base and height of the GTP to be equal to $p$. In this case, it is straightforward to show that the number of edges on each of the two sides remains equal, this number being $k={\rm gcd}(p,q)={\rm gcd}(p,p-q)$. We can express $(p,q)=k\,(P,Q)$, with $P$ and $Q$ coprime. Let us denote $V_k$ the T-cone that they define. Computing the self-intersection, we obtain $\mathcal{I}=-2k^2$. Therefore, primitive T-cones saturate the SUSY bound while those for $k>1$ appear to be non-SUSY. This is merely an artifact of $V_k$  being a superposition of $k$ copies of the same web, and hence do not correspond to an irreducible curve for $k>1$ \cite{Bergman:2020myx}. It is clear that $V_k$ is the superposition of $k$ copies of $V_1$, and so it is SUSY.\footnote{One can note that, forgetting about the coloring of dots in $V_k$, it can be generated as the Minkowski sum $V_k=\bigoplus_{i=1}^k V_1$.}

\section{T-cones, locked superpositions and the extended Coulomb branch} \label{TconesGTP}

Let us consider a brane web realizing a rank $r$ $5d$ SCFT with $g$ mass deformations. This means that the web admits $r$ local deformations (opening of faces) and $g$ global deformations (changing the asymptotic positions of the external legs). Let us start by discussing webs in which all legs terminate on different 7-branes, i.e. webs associated with ordinary toric diagrams. In this case, a triangulation of the toric diagram by minimal area 1/2 triangles, which in the language of this paper correspond to $p=1$ T-cones, contains  $n_T=2r+g+1$ triangles, following Pick's theorem. Conversely, engineering the $5d$ SCFT through M-theory on a $CY_3$, the extended Coulomb branch arises as the set of all possible resolutions of the $CY_3$.

Now, let us consider webs with non-trivial external multiplicities, namely webs associated to general GTPs. In this case, to reveal the Coulomb branch, we must `open up’ the web as much as possible by separating all allowable triple junctions with self-intersection $\mathcal{I} = -2$, ensuring that no faces remain hidden \cite{Bergman:2020myx}.\footnote{Below, in Section \sref{locked}, we will see that this is not fully general. In the physics context, it is well-known that the S-rule can sometimes enforce configurations where the web opens such that 5-branes pass over other 5-branes without breaking \cite{Benini:2009gi}. For now, let us focus on webs that do not require such jumps and defer a discussion of the generic case to Section \sref{locked}.} As we have seen, such junctions correspond to T-cones. Therefore, in the generic case, moving into the extended Coulomb branch corresponds, geometrically, to desingularizing the associated $CY_3$ through resolutions and local $\mathbb{Q}$-Gorenstein smoothings, the latter accounting for the presence of 7-branes. We naturally arrive at the following recipe for constructing a generic point in the extended Coulomb branch of a $5d$ theory engineered by a brane web: starting from a GTP with external dots colored according to how the legs of the web terminate on 7-branes, we tessellate its interior with T-cones, determining in the process the color of internal dots. This algorithm recovers the physical prescription (see e.g. \cite{Bergman:2020myx}). Denoting the set of all T-cones as $\mathcal{T}$, we can regard the extended Coulomb branch as a puzzle whose pieces are the elements of $\mathcal{T}$ and with a boundary dictated by the GTP. In the coming section, we will see that the set of pieces needs to be enlarged beyond T-cones.

The idea of capturing the extended Coulomb branch by tessellations of the GTP is not new. It was indeed introduced in the original work that initiated this line of research \cite{Benini:2009gi}. However, it is worth noting that the elementary pieces we use to tile GTPs differ from the ones in \cite{Benini:2009gi}, which did not contemplate T-cones and, hence, their importance.

\subsection{The most general pieces: locked superpositions}\label{locked}

We have argued that primitive T-cones are elementary building blocks for tessellating GTPs. Their corresponding webs are triple intersections without mass deformations or Coulomb branch. From this perspective, these webs stand out in that they cannot be further opened up, as they lack additional degrees of freedom to span an extended Coulomb branch. This suggests that the set of building blocks for tessellating GTPs should be expanded to include all possible webs corresponding to 5D SCFTs without an extended Coulomb branch. Clearly, in addition to primitive T-cones, these also include generic, non-primitive T-cones. However, as described in Section \ref{noprim}, the latter can be viewed as superpositions of primitive T-cones.

\begin{figure}[h!]
\centering
\includegraphics[width=\textwidth]{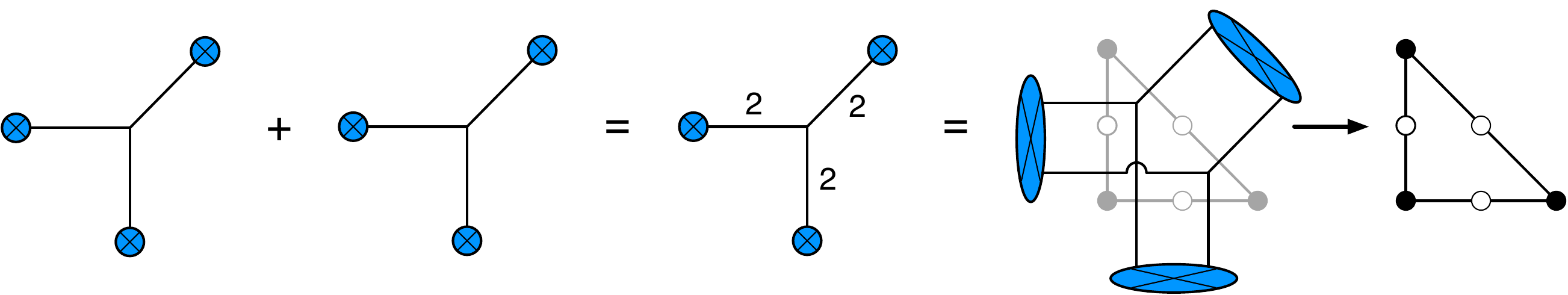}
\caption{A locked superposition of two webs corresponding to primitive $(1,0)$ T-cones. The result is a non-primitive $(2,2)$ T-cone.}
\label{locked 1}
\end{figure}

\begin{figure}[h!]
\centering
\includegraphics[width=.97\textwidth]{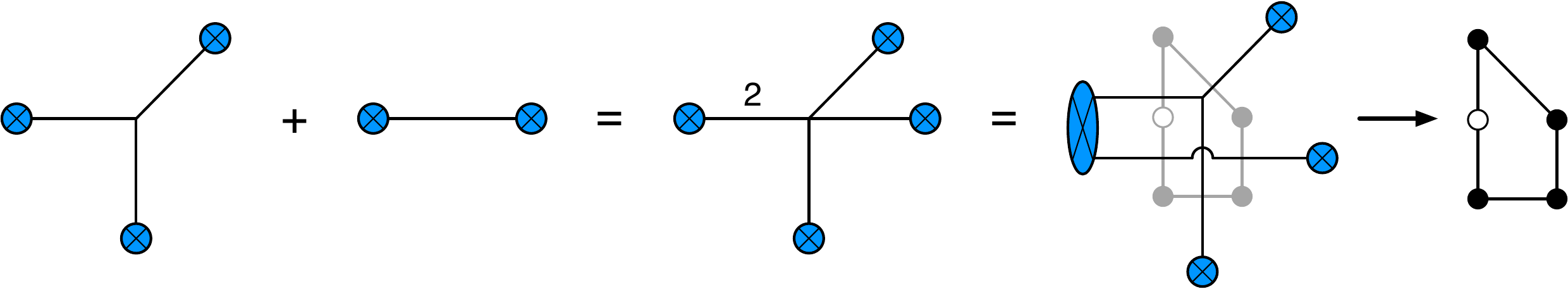}
\caption{Another locked superposition of webs. The result is a $5d$ theory with no extended Coulomb branch.}
\label{locked 2}
\end{figure}

We can use this observation to generalize the primitive T-cones and construct all possible fundamental tiles. Focusing on the global symmetry, a web with $L$ external legs that are free to move independently has a rank $L-3$ global symmetry. However, some of these degrees of freedom can be 'locked' by terminating multiple subwebs on the same 7-branes, constraining the movement of their legs so they are no longer independent. We will refer to such configurations as \textit{locked superpositions}. Note that for locked configurations, the supersymmetry formula in terms of the self-intersection of the curve does not work \cite{Bergman:2020myx}, as the web has disconnected components. Non-primitive T-cones can be regarded as locked superpositions of primitive T-cones. The most general tiles satisfying the criteria of not having an extended Coulomb branch correspond to locked superpositions of T-cones and single 5-branes. The examples in Figures \ref{locked 1} and \ref{locked 2} illustrate this approach for constructing theories with no extended Coluomb branch.

In summary, we have to extend the set of pieces $\mathcal{T}$ for tessellating GTPs, which becomes:
\begin{enumerate}
\item Triple intersections corresponding to primitive T-cones.
\item Locked superpositions of primitive T-cones and trivial webs (i.e. brane webs composed of a 5-brane segments). These configurations include the triangles and trapezia introduced in  \cite{Benini:2009gi}.
\end{enumerate}
These rules go beyond \cite{Benini:2009gi}, as shown in \fref{GTP - Web -  possible tiles}

\begin{figure}[h!]
\centering
\includegraphics[height=7cm]{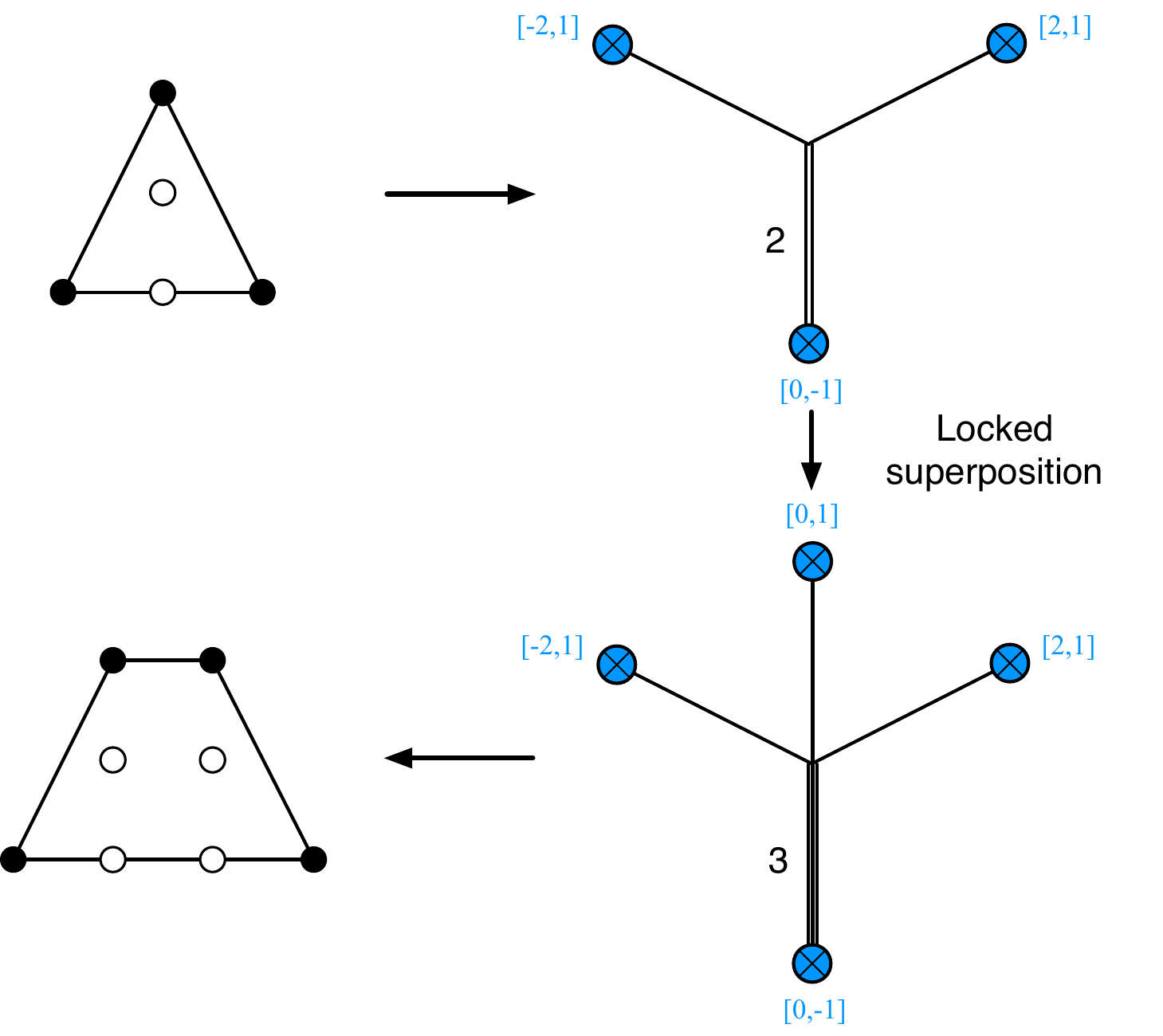}
\caption{Minimal polygons allowed in tessellations of GTPs that go beyond the rules in \cite{Benini:2009gi}.}
\label{GTP - Web - possible tiles}
\end{figure}

We refer to these tessellations as {\it tangrams}, due to their similarity to the namesake puzzle, which involves tiling various figures using differently shaped pieces.

\subsection{Different tessellations}

\begin{figure}[h!]
\centering
\includegraphics[height=7.5cm]{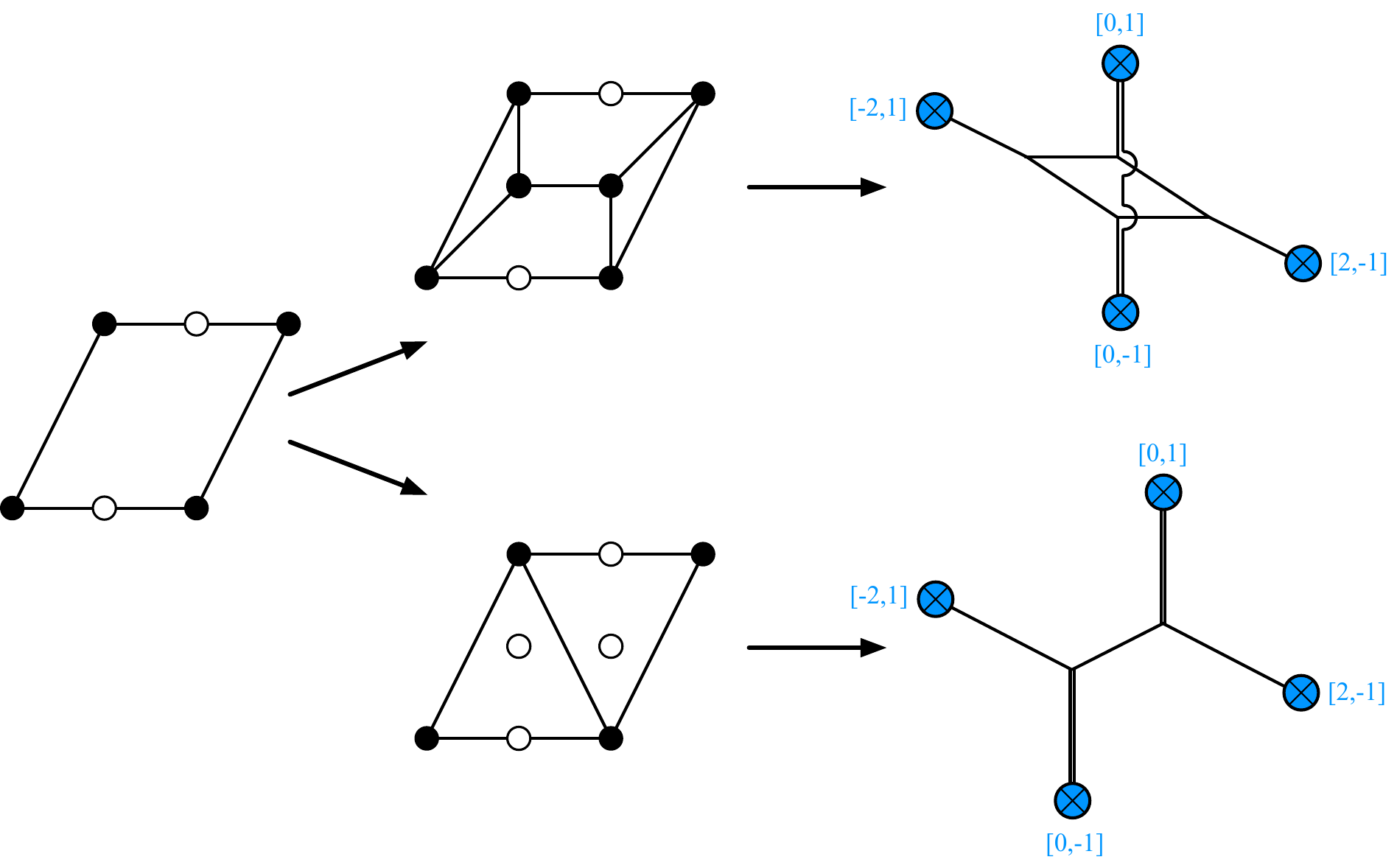}
\caption{Tessellations of a GTP containing different numbers of elementary pieces.}
\label{Different tessellations}
\end{figure}

Once we allow the set of elementary pieces to include generic T-cones and locked superpositions, a natural question is whether all possible tessellations of a given GTP are equivalent. More fundamentally, do all tessellations contain the same number of pieces? It is clear that all triangulations of ordinary toric diagrams, which consist of minimal area 1/2 triangles, have the same number of pieces. But this is no longer obvious once we consider pieces with different areas. It is straightforward to conclude GTPs may admit tessellations with different numbers of minimal pieces, as in the examples shown in \fref{Different tessellations}.

We can understand what is going on by inspecting the corresponding webs. The tessellation at the top corresponds to motion in the Coulomb branch, while the lower tessellation corresponds to a mass deformation (i.e. motion in the extended Coulomb branch directions transverse to the Coulomb branch), which is not a gauge theory. Clearly, to connect both webs it is necessary to pass through the origin. We conclude that the extended Coulomb branch is the union of two cones touching at the origin: one corresponding to the Coulomb branch, the other to the mass deformation. This is a very interesting phenomenon that deserves further study.

\section{Algebraic deformation as a deformation of the BPS quiver: an explicit example}\label{explicit}

In the particular case of webs whose external legs are all single five-branes the associated GTP is an actual toric diagram corresponding to that of the toric CY$_3$ which geometrically engineers the singularity. This correspondence allows the construction of the BPS quiver, which, as described in \cite{Closset:2019juk}, coincides with the fractional brane quiver for 3-branes in Type IIB String Theory probing the $CY_3$ singularity. This allows to import the heavy machinery developed to construct and study such theories (\textit{e.g.} \cite{Feng:2000mi,Franco:2005sm,Franco:2005rj}). The BPS quiver provides then a tool to explore the geometric engineering of 5d SCFT's, including Hanany-Witten transitions where a 7-brane is moved across the web. In this section, we present a concrete example where the superpotential deformations of the BPS quiver between two theories introduced in \cite{Cremonesi:2023psg} appear as algebro-geometric one-parameter deformations of the varieties defining each theory. These are parameterized by $t$, with the generic fiber ($t\neq 0$) and the exceptional fiber ($t=0$) corresponding to the geometries of the deformed and undeformed theories, respectively. This explicitly realizes the construction proposed in \cite{Ilten_2012} at the level of the BPS quiver, in turn corresponding to the HW transition, as conjectured in \cite{Arias-Tamargo:2024fjt}.

\begin{figure}[h!]
\centering
\includegraphics[height=5.5cm]{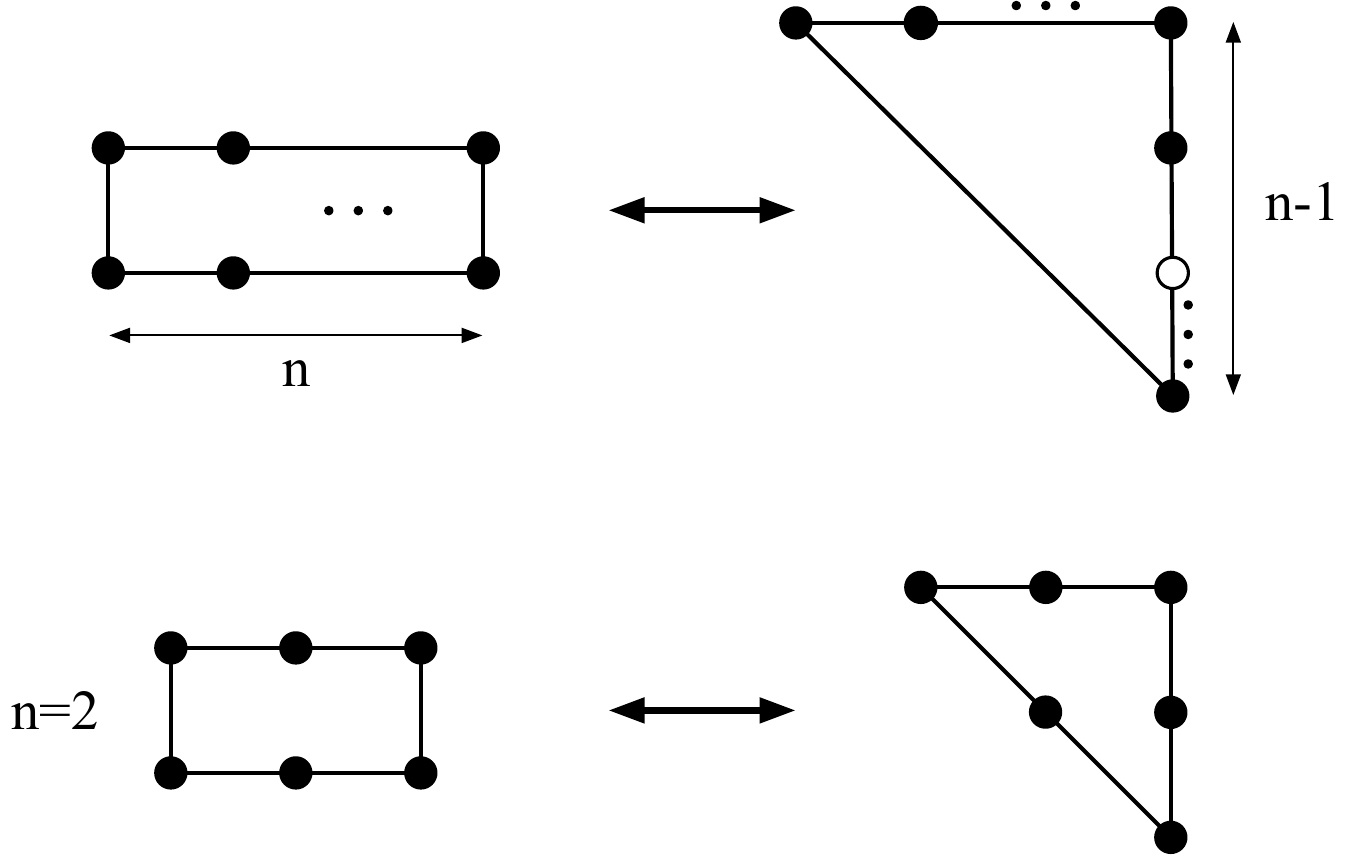}
\caption{A family of GTPs and their mutations.}
\label{GTPs mutations}
\end{figure}

\begin{figure}[h!]
\centering
\includegraphics[height=5.5cm]{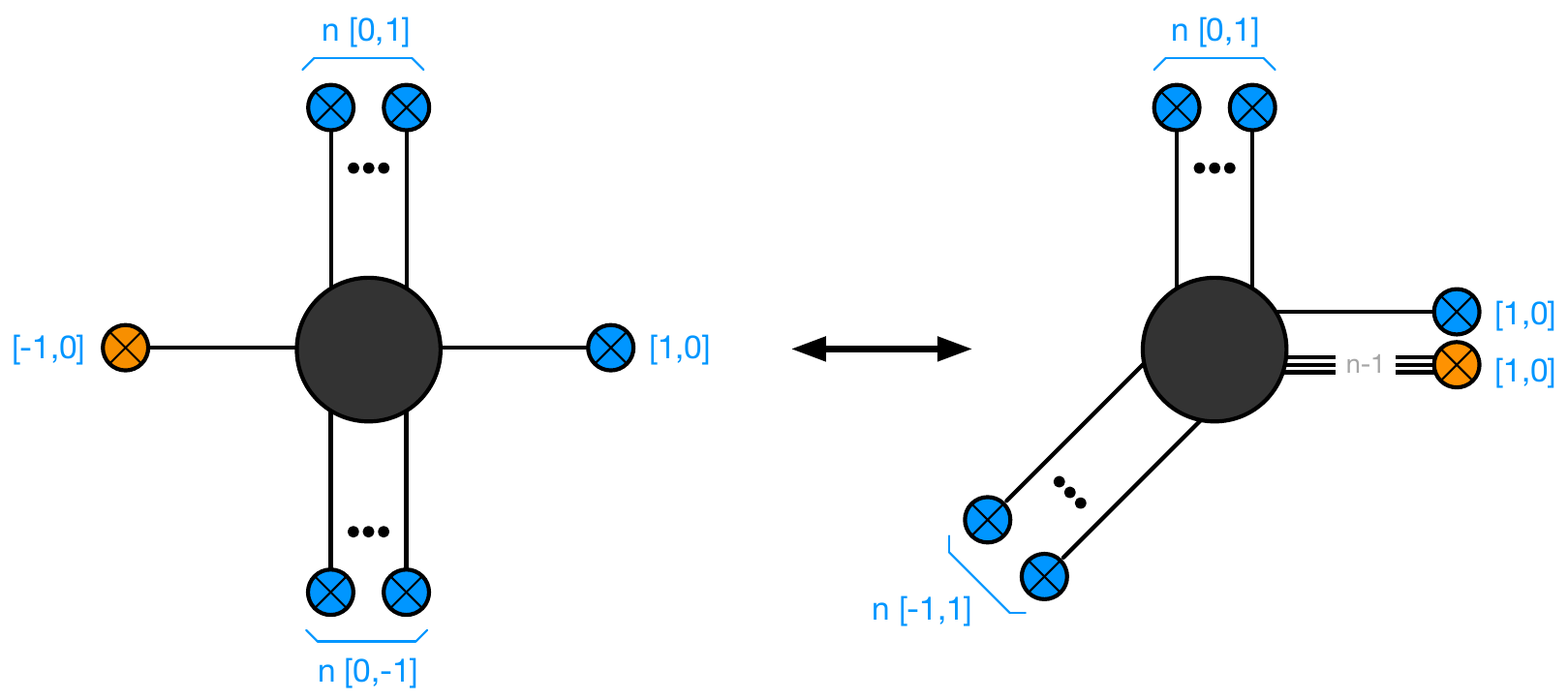}
\caption{Brane webs for the GTPs in \fref{GTPs mutations}.}
\label{web_conifoldZN_and_mutation}
\end{figure}

Consider the family of theories represented by the toric diagrams in the upper left corner of \fref{GTPs mutations}, corresponding to non-chiral $\mathbb{Z}_n$ orbifolds of the conifold. The associated webs are shown in \fref{web_conifoldZN_and_mutation}. Crossing the left brane to the opposite side yields a brane web represented by the GTP in the upper right corner of \fref{GTPs mutations}, with an underlying toric diagram corresponding to $\mathbb{C}^3/\mathbb{Z}_n\times \mathbb{Z}_n$. We present details of the algebraic characterization of these varieties in Appendix \sref{varieties}.

As discussed in \cite{Akhtar_2012}, the Hilbert series is an invariant under mutation. Invariance under a transformation that alters the underlying GTP might come as a surprise to those familiar with the computation of Hilbert series from quivers.  It turns out that the Hilbert series under consideration, and hence its invariance, depends on the choice of gradings for fields.\footnote{Very recently \cite{Bao:2024nyu} also considered different gradings in the Hilbert series. It would be very interesting to study the choice in this paper in those terms.} The Hilbert series is obviously not invariant for the grading chosen in Appendix \sref{varieties} (cf. \eqref{HSConZn} and \eqref{HSC3ZnZn}). However, this can be fixed by assigning gradings to the algebraic generators in the Hilbert series that are different from 1. Let $R_1$, $R_2$, $R_3$ and $R_4$ be the gradings of the four generators of Conifold/$\mathbb{Z}_n$ and $\mathbb{C}^3/\mathbb{Z}_n \times \mathbb{Z}_n$. From their defining equations, we impose
\begin{equation}
n(R_1+R_2)=R_3+R_4\,,\quad n R_4=R_1+R_2+R_3 \, .
\end{equation}
Both equations have the same grading. This choice is achieved by declaring the following gradings for the coordinates
\begin{equation}
{\rm conifold}/\mathbb{Z}_n: \qquad \begin{array}{|c|c|c|c|c|c} \hline  
& u & v & w & t \\  \hline 
\ {\rm Grading} \ \ & \ R_1 \ \ & \ R_2 \ \ & \ \frac{1}{n}(R_1+R_2) \ \ & \ \frac{n-1}{n}(R_1+R_2) \ \ \\ \hline
\end{array}
\end{equation}
for ${\rm conifold}/\mathbb{Z}_n$ in the presentation of \eqref{conifoldmodZn} and
\begin{equation}
\mathbb{C}^3/\mathbb{Z}_n\times \mathbb{Z}_n: \qquad  \begin{array}{ |c|c|c|c|} \hline
  & x & y & z \\  \hline 
\ {\rm Grading} \ \ & \ \frac{(n-1)}{n}\,(R_1+R_2) \ \ & \ \frac{R_1}{n} \ \ & \ \frac{R_2}{n} \ \ \\ \hline
\end{array}
\end{equation}
for $\mathbb{C}^3/\mathbb{Z}_n\times \mathbb{Z}_n$ in the presentation of \eqref{C3modZnZn}. The, one finds in both cases
\begin{equation}
\label{HS}
HS_{{\rm conifold}/\mathbb{Z}_n}=HS_{\mathbb{C}^3/\mathbb{Z}_n^2}={\rm PE}[t^{R_1+R_2}+t^{(n-1)(R_1+R_2)}+t^{R_1}+t^{R_2}-t^{n(R_1+R_2)}]\,.
\end{equation}

\subsection{The case of $n=2$}

\begin{figure}[H]
\centering
\includegraphics[height=5cm]{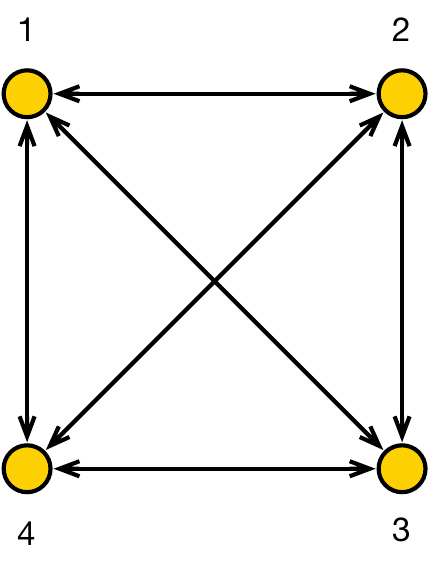}
\caption{Quiver diagram for $\mathbb{C}^3/(\mathbb{Z}_2 \times \mathbb{Z}_2)$}
\label{quiver_C3_Z2xZ2}
\end{figure}

The simplest example in this family is the case $n=2$, which, while not corresponding to an actual GTP, serves well for illustrative purposes. Taking as starting point $\mathbb{C}^3/\mathbb{Z}_2\times \mathbb{Z}_2$, the BPS quiver is shown in \fref{quiver_C3_Z2xZ2}.

The superpotential for this theory is
\begin{eqnarray} 
W&=& X_{42}X_{23}X_{34}+X_{43}X_{31}X_{14}+X_{41}X_{12}X_{24}+X_{13}X_{32}X_{21} \nonumber \\ 
 & - & X_{12}X_{23}X_{31}-X_{14}X_{42}X_{21}-X_{34}X_{41}X_{13}-X_{43}X_{32}X_{24}\,.
 \end{eqnarray}

Assuming the following gradings for fields
\begin{equation}
\{X_{12},X_{21},X_{23},X_{32}X_{34},X_{43},X_{14},X_{41}\}\rightarrow 1\,,\qquad \{X_{24},X_{42},X_{13},X_{31}\}\rightarrow 2\,,
\end{equation}
we obtain exactly the expected result in \eqref{HS} with $R_1=R_2=2$. 

Let us now deform the superpotential by turning on the following deformation
\begin{equation}
\delta W=\mu\,(X_{24}X_{42}-X_{13}X_{31})\,.
\end{equation}
Note that the deformation is homogeneous since it has grading 4 like the rest of the superpotential.

\begin{figure}[H]
\centering
\includegraphics[height=5cm]{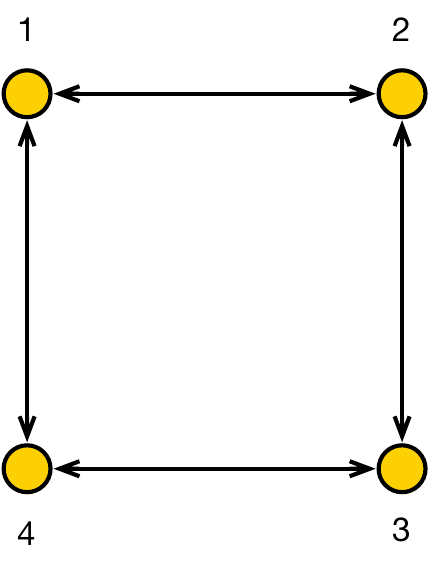}
\caption{Quiver diagram for the $\mathbb{Z}_2$ orbifold of the conifold.}
\label{quiver_conifold_Z2}
\end{figure}

Integrating out massive fields, we obtain the quiver in \fref{quiver_conifold_Z2}, with superpotential 
\begin{eqnarray}
\nonumber W&=& -X_{21}X_{12}X_{23}X_{32}+X_{12}X_{21}X_{14}X_{41}\\ 
& & -  X_{41}X_{14}X_{43}X_{34}+X_{32}X_{23}X_{34}X_{43}\,,
\end{eqnarray}
which corresponds to the ${\rm Conifold}/\mathbb{Z}_2$,.

Assuming now equal grading for all fields, we generate again the expected result in \eref{HS} with $R_1=R_2=2$.\footnote{From now on, we leave possible overall rescalings of the $t$ variable implicit.}

To understand the deformation in terms of algebraic geometry, let us return to the undeformed quiver in \fref{quiver_C3_Z2xZ2}. Upon using the relations coming from vanishing $F$-terms, we can consider the following independent gauge invariant operators
\begin{equation}
A=X_{23}X_{34}X_{42}\,,\qquad B=X_{42}X_{24}\,,\qquad C=X_{14}X_{41}\,,\qquad D=X_{34}X_{43} \, .
\end{equation}
Using the $F$-terms, these satisfy $A^2=BCD$. Note that this leads to $[A]=4$, $[B]=4$, $[C]=[D]=2$ with $[W]=4$. This suggests we can solve the equation by introducing three variables $\{\tilde{x},\,y,\,z\}$ with equal grading as follows
\begin{equation}
\label{par2}
A=\tilde{x}^2yz\,,\qquad b=\tilde{x}^4\,,\qquad c=y^2\,,\qquad d=z^2\,.
\end{equation}
This coincides with \eqref{scal1} for $n=2$ upon setting $x=\tilde{x}^2$. The non-standard powers derived above suggest that, while the defining equation of the geometry remains the same, the resulting variety differs slightly. The map between these varieties includes branch cuts, suggesting the variety is not fully covered. This is reminiscent of how a set of algebraic equations (a toric set) may only partially define a toric variety (see for example exercise 1.1.6 in \cite{CLS} for more details).

Turning on the deformation, the independent gauge invariants modulo $F$-terms remain the same, but the equation becomes
\begin{equation}
\label{eqdeformed}
A^2-BCD+\mu\,AB=0\,.
\end{equation}
Let us introduce
\begin{equation}
A=a-\frac{\mu}{2}b\,,\qquad B=b,\,\qquad C=c\,,\qquad\epsilon=\frac{\mu^2}{4}\, .
\end{equation}
In terms them, the deformed equation becomes
\begin{equation}
\label{def}
a^2-bcd-\epsilon b^2=0\,.
\end{equation}
This is precisely the equation defining the $\mathbb{Q}$-Gorenstein smoothing of $\mathbb{C}^3/\mathbb{Z}_2\times \mathbb{Z}_2$ as described in 7.3 of \cite{Biquard-Rollin}. To see this explicitly, we can solve \eqref{def} by setting
\begin{equation}
\label{sol}
a=\tilde{a}-\frac{\tilde{b}}{4}\,,\qquad  b=-\frac{1}{\epsilon^{\frac{1}{2}}}\big(\tilde{a}+\frac{\tilde{b}}{4}+\tilde{c}\,\tilde{d}\big)\,,\qquad c=(4\,\epsilon)^{\frac{1}{4}}\,\tilde{c}\,,\qquad d=(4\, \epsilon)^{\frac{1}{4}}\,\,\tilde{d}\,,
\end{equation}
provided $\tilde{a}\tilde{b}-\tilde{c}^2\tilde{d}^2=0$. Thus we have that for any $\epsilon\ne 0$ the geometry is ${\rm conifold}/\mathbb{Z}_2$, while for $\epsilon=0$ it is that of $\mathbb{C}^3/\mathbb{Z}_2\times \mathbb{Z}_2$.

It is interesting to come back to the issue of the gradings. We have found that the deformation of the BPS quiver corresponds to the smoothing of $\mathbb{C}^2/\mathbb{Z}_2\times \mathbb{Z}_2$ into ${\rm conifold}/\mathbb{Z}_2$ described by
\begin{equation}
a^2=bcd+\epsilon\,b^2\,.
\end{equation}
This is shown explicitly by \eref{sol}, which expresses ${a, b, c, d}$ in terms of ${\tilde{a}, \tilde{b}, \tilde{c}, \tilde{d}}$. The latter parametrize the ${\rm conifold}/\mathbb{Z}_2$, since they satisfy $\tilde{a}\tilde{b}=\tilde{c}^2\tilde{d}^2$. Let us suppose that we describe such variety in the standard way, as discussed in Appendix \sref{appendix_geometry}. This naturally assigns gradings $[\tilde{a}]=\tilde[b]=2R$, $[\tilde{c}]=[\tilde{d}]=R$ (\textit{cf.} \eqref{scal1}). Assuming $\epsilon$ to be a dimensionless parameter, we find from \eref{sol} that $[a] = [b] = 2R$ and $[c] = [d] = R$. Comparing this with \eref{par}, it is evident that these are not the natural gradings. Now, consider the case $\epsilon = 0$ and insist on these gradings; that is, we attempt to solve $a^2 = bcd$ with $[a] = [b] = 2R$ and $[c] = [d] = R$. This can be attained, for instance, by the change of variables $x\rightarrow xy$ in \eref{par}, that is
\begin{equation}
\label{par2}
a=xy^2z\,,\qquad b=x^2y^2\,,\qquad c=y^2\,,\qquad d=z^2\,.
\end{equation}
While this parametrization solves the same equation, it globally describes a different variety. To see this, note that in the parametrization of \eref{par}, the origin comprises the union of three cones where $b = 0$, $c = 0$, or $d = 0$. By contrast, in the parametrization of \eref{par2}, the origin corresponds to the union of the cones $b = 0$, $b = c = 0$, and $d = 0$.

\appendix

\section{Some details on the geometry of conifold$/\mathbb{Z}_n$ and $\mathbb{C}^3/\mathbb{Z}_n\times \mathbb{Z}_n$ }\label{varieties}

\label{appendix_geometry}

This appendix provides details on the construction of the varieties ${\rm conifold}/\mathbb{Z}_n$ and $\mathbb{C}^3/\mathbb{Z}_n \times \mathbb{Z}_n$, along with their Hilbert series.

\subsection{Conifold$/\mathbb{Z}_n$}

The ${\rm conifold}/\mathbb{Z}_n$ variety can be characterized as
\begin{equation}
\label{conifoldmodZn}
{\rm conifold}/\mathbb{Z}_n = \left\{ \frac{(u,y,w,t)}{uv=wt,\,(w,t)\sim(\omega w,\omega^{-1}t)}\right\}\,,\qquad \omega^n=1\,.
\end{equation}
Alternatively, we can consider the invariants under the $\mathbb{Z}_n$ action
\begin{equation}
a=w^n\,,\qquad b=t^n\,,\qquad c=u\,,\qquad d=v\,. \label{scal1}
\end{equation}
Note that we could also have chosen the $w$ and $t$ variables but, through $uv=wt$, this is equivalent to our choice. The invariants satisfy $ab=c^n d^n$, so we can write
\begin{equation}
{\rm conifold}/\mathbb{Z}_n = \left\{ \frac{(a,b,c,d)}{ab=c^nd^n}\right\}\,.
\end{equation}

We may double-check this description by computing the Hilbert series (see \textit{e.g.} \cite{benvenuti2007counting}). Starting with the description in terms of $\{u,v,w,t\}$ and introducing a fugacity $\alpha$ for the $U(1)$ acting on $(w,t)$, we use the plethystic exponential to determine
\begin{equation}
HS_{\rm conifold}(\alpha)={\rm PE}[2t+t(\alpha+\alpha^{-1})-t^2]=\frac{(1-t^2)}{(1-t)^2\,(1-t\alpha)\,(1-t\alpha^{-1})}\,.
\end{equation}
Quotienting by $\mathbb{Z}_n$, we find
\begin{equation}
HS_{\rm {\rm conifold}/\mathbb{Z}_n}=\frac{1}{n}\,\sum_{k=0}^{n-1}HS_{\rm conifold}(e^{i\frac{2\pi}{n}k})={\rm PE}[2t+2t^n-t^{2n}]\,, \label{HSConZn}
\end{equation}
which is precisely the description in terms of the invariants $\{a,b,c,d\}$. By taking the plethystic logarithm of \eqref{HSConZn} we can extract information about the generators and their relations. We see then that the invariants $c$ and $d$ have grading 1, the invariants $a$ and $b$ have grading $n$, and they are related by an equation of grading $2n$.

\subsection{$\mathbb{C}^3/\mathbb{Z}_n\times \mathbb{Z}_n$}

Let us now consider $\mathbb{C}^3/\mathbb{Z}_n\times \mathbb{Z}_n$. It can be described as
\begin{equation}
\label{C3modZnZn}
\mathbb{C}^3/\mathbb{Z}_n\times \mathbb{Z}_n=\{(x,y,z)/\sim\}\,,\qquad \sim:=\, \begin{cases} (x,y,z)\sim (\omega_1 x,\omega_1^{-1}y,z)\\  (x,y,z)\sim ( x,\omega_2 y,\omega_2^{-1}z)\end{cases}
\end{equation}
where $\omega_i^n=1$. The invariants under the orbifold action are
\begin{equation}
\label{par}
a=xyz\,,\qquad b=x^n\,,\qquad c=y^n\,,\qquad d=z^n\,.
\end{equation}
Therefore, we can regard the variety as a hypersurface in $\mathbb{C}^4$ defined as
\begin{equation}
\mathbb{C}^3/\mathbb{Z}_n^2=\left\{\frac{(a,b,c,d)}{a^n=bcd}\right\}\,.
\end{equation}
We can again check this descriptionusing the Hilbert series. In terms of the $\{x,y,z\}$, including the fugacities $\alpha$ and $\beta$ for each of the $\mathbb{Z}_n$ actions, and plethystic exponential, the refined Hilbert series for $\mathbb{C}^3$ is 
\begin{equation}
HS_{\mathbb{C}^3}(\alpha,\beta)=\frac{1}{(1-t\alpha)\,(1-t\alpha^{-1}\beta)\,(1-t\beta^{-1})}\,.
\end{equation}
Then, implementing the $\mathbb{Z}_n\times \mathbb{Z}_n$ quotient gives
\begin{equation}
HS_{\mathbb{C}^3/\mathbb{Z}_n^2}=\frac{1}{n}\sum_{k_1=0}^{n-1}\,\frac{1}{n}\sum_{k_2=0}^{n-1}HS_{\mathbb{C}^3}(e^{i\frac{2\pi}{n}k_1},e^{i\frac{2\pi}{n}k_2})={\rm PE}[t^3+3t^n-t^{3n}]\,, \label{HSC3ZnZn}
\end{equation}
which recovers the description in terms of $\{a,b,c,d\}$. The invariant $a$ has grading $3$, the invariants $b$, $c$ and $d$ have grading $n$, and there is one equation of grading $3n$ relating them.

\section{Self-intersection formula}\label{SI}

An important quantity related to a GTP is the self-intersection $\mathcal{I}$ \cite{Bergman:2020myx}. Let $L_i=\{(p_i,q_i)\}$ be the $(p,q)$ charges of the external 7-branes, arranged in clockwise order. The self-intersection is defined as
\begin{equation}
\mathcal{I}= \left|\sum_{i<j} \det \left(\begin{array}{cc}
p_i&q_i\\
p_j&q_j
\end{array}\right)\right|-\sum_i \gcd(p_i,q_i)^2\,. \label{selfint}
\end{equation}

The rank of the theory defined by the GTP is simply $r=\frac{\mathcal{I}+2}{2}$. If the GTP is a toric diagram (i.e. if  $\gcd(p_i,q_i)=1$ for all $i$), the rank equals the number of interior points. We remark that this formula is only valid if the underlying brane web is connected, in the sense that subwebs cannot be slid along the directions of the 7-branes.

A GTP whose rank is greater than zero can be opened up. This means taking a junction (understood as a GTP by itself) with $r>0$ and replacing it with a subweb composed of $r$ faces, whose junctions are all triple with $\mathcal{I}=-2$. From a gauge theory point of view, this corresponds to ''turning on'' $r$ Cartans.

\acknowledgments

We would like to thank Guillermo Arias-Tamargo, Francesco Benini, Sergio Benvenuti, Antoine Bourget, Julius Grimminger, Amihay Hanany and Francesco Mignosa for discussions. We would like to acknowledge CERN, the Simons Physics Summer Workshop and the Simons Center for Geometry and Physics for their hospitality during part of this work. S.F. is supported by the U.S. National Science Foundation grants PHY-2112729 and PHY-2412479.  I.C.B and D.R.G are supported in part by the Spanish national grant MCIU-22-PID2021-123021NB-I00.


\bibliographystyle{JHEP}
\bibliography{mybib}

\end{document}